\def\rfr#1{Eq. (\ref{#1})}

\def\derp#1#2{\rp{\partial{#1}}{\partial{#2}}}

\def\bar{\begin{eqnarray}}
\def\ear{\end{eqnarray}}

\def\eqi{\begin{equation}}
\def\eqf{\end{equation}}
\def\eqia{\begin{eqnarray}}
\def\eqfa{\end{eqnarray}}
\def\rp#1#2{{#1\over#2}}
\def\ct#1{\cite{#1}}
\def\lb#1{\label{#1}}

\def\oc2{$\mathcal{O}(c^{-2})$}

\documentclass[11pt]{article}
\usepackage{amsmath,amsthm,amscd,amssymb}
\usepackage{latexsym}
\usepackage{graphicx,epsfig,lscape,rotating}

\begin{document}

\noindent{\bf \LARGE{On the mean anomaly and the Lense-Thirring effect }}
\\
\\
\\
{L. Iorio, \textit{FRAS, DDG} }\\
{\it Viale Unit$\grave{a}$ di Italia 68, 70125\\Bari, Italy
\\tel./fax 0039 080 5443144
\\e-mail: lorenzo.iorio@libero.it}

\begin{abstract}
In this brief note we reply to the authors of a recent preprint in
which an alleged explicit proposal of using the mean anomaly of
the LAGEOS satellites to measure the general relativistic
Lense-Thirring effect in the gravitational field of the Earth is
attributed to the present author.
\end{abstract}



The authors of the recent preprint \cite{cp06}, posted on 4th
January 2006, claim that the present author would have explicitly
proposed to use the mean anomaly of the LAGEOS satellites for
increasing the precision of the measurements of the Lense-Thirring
effect in the gravitational field of the Earth. They support their
claim with the following citation from two old, unpublished
versions of the preprints \cite{ior04a} (19th April 2005) and
\cite{ior04b} (19th April 2005): ``The problem of reducing the
impact of the mismodeling in the even zonal harmonics of the
geopotential with the currently existing satellites can be coped
in the following way.

Let us suppose we have at our disposal N (N$>1$) time series of
the residuals of those Keplerian orbital elements which are
affected by the geopotential with secular precessions, i.e. the
node, the perigee and the mean anomaly: let them be $\psi^{\rm
A},$ A=LAGEOS, LAGEOS II, etc. Let us write explicitly down the
expressions of the observed residuals of the rates of those
elements $\delta\dot\psi^{\rm A}_{\rm obs}$ in terms of the
Lense-Thirring effect $\dot\psi_{\rm LT}^{\rm A}$, of N-1
mismodelled classical secular precessions $\dot\psi_{.\ell}^{\rm
A}\delta J_{\ell}$ induced by those even zonal harmonics whose
impact on the measurement of the gravitomagnetic effect is to be
reduced and of the remaining mismodelled phenomena $\Delta$ which
affect the chosen orbital element \eqi\delta\dot\psi_{\rm
obs}^{\rm A}=\dot\psi_{\rm LT}^{\rm A}\mu_{\rm LT }+\underset{{\rm
N-1\ terms }}{\sum}\dot\psi_{.\ell}^{\rm A }\delta
J_{\ell}+\Delta^{\rm A},\ \underset{{\rm N}}{\underbrace{{\rm
A=LAGEOS,\ LAGEOS\ II,...}}} \lb{equaz}\eqf The parameter
$\mu_{\rm LT}$ is equal to 1 in the General Theory of Relativity
and 0 in Newtonian mechanics. The coefficients
$\dot\psi_{.\ell}^{\rm A}$ are defined as
\eqi\dot\psi_{.\ell}=\derp{{\dot\psi}_{\rm class}}{J_{\ell}}\eqf
and have been explicitly worked out for the node and the perigee
up to degree $\ell=20$ in Iorio (2002b; 2003a); they depend on
some physical parameters of the central mass ($GM$ and the mean
equatorial radius $R$) and on the satellite's semimajor axis $a$,
the eccentricity $e$ and the inclination $i$. We can think about
\rfr{equaz} as an algebraic nonhomogeneuous linear system of N
equations in N unknowns which are $\mu_{\rm LT}$ and the N-1
$\delta J_{\ell}$: solving it with respect to $\mu_{\rm LT}$
allows to obtain a linear combination of orbital residuals which
is independent of the chosen N-1 even zonal harmonics.''.

The authors of \cite{cp06} seem to be not aware of  \cite{ior05},
posted on 3rd August 2005, in which this question was already
tackled and fully explained (Section 4, pag.7). Thus, we invite
the interested readers and the authors of \cite{cp06} to go
through \cite{ior05}. In it there are also detailed analyses
(Section 2) of the unfeasible proposal by the authors of
\cite{cp06} of using the existing polar satellites to measure the
Lense-Thirring effect and a quantitative discussion of the
possibility of using the satellites Jason-1 and Ajisai (Section
3).

Here we limit to note that the interpretation of the cited passage
put forth by the authors of \cite{cp06} is misleading and
untenable. Indeed,  in \ct{ior04a, ior04b}, and other papers of
him, the author of this note presented in the most general way the
linear combination approach simply enumerating the Keplerian
orbital elements affected by the even zonal harmonics of the
Earth's geopotential with secular precessions, i.e. the node, the
perigee and the mean anomaly. In no way that fact can be assumed
as an explicit proposal of using the mean anomaly for testing the
Lense-Thirring effect: for example, no explicit linear
combinations including such an orbital element can be found in all
the published (and unpublished) works by the present author.
Moreover, in \cite{ior04a, ior04b} it is explicitly written just
after the passage previously cited by the authors of
\cite{cp06}:``In general, the orbital elements employed are the
nodes and the perigees [...]'' (Section 2.1.2, pag.8 of
\cite{ior04a} and Section 2.1.1, pag.6 of \cite{ior04b}). The
authors of \cite{cp06} miss to cite such statement.

As already pointed out in \cite{ior05} (Section 4, pag.7),  one of
the authors of \cite{cp06} did explicitly use the mean anomaly of
LAGEOS II in some tests conducted with the EGM96 model and
published in papers and proceedings books.


\end{document}